\documentclass[12pt]{article}
\usepackage{graphicx,amsmath}
\usepackage{units}

\newlength{\dinwidth}
\newlength{\dinmargin}
\def\lapproxeq{\lower .7ex\hbox{$\;\stackrel{\textstyle                                                    
<}{\sim}\;$}}                                                    
\def\gapproxeq{\lower .7ex\hbox{$\;\stackrel{\textstyle                                                    
>}{\sim}\;$}}                                                    
\def\be{\begin{equation}}                                                    
\def\ee{\end{equation}}                                                    
\def\bea{\begin{eqnarray}}                                                    
\def\eea{\end{eqnarray}}

\def\GeV{\rm GeV}

\def\sh{\hat s}
\def\sh2{{\hat s}^2}

\begin{document}

\begin{flushright}                                                    
IPPP/12/07  \\
DCPT/12/14 \\                                                    
\today \\                                                    
\end{flushright} 

\vspace*{0.5cm}

\begin{center}
{\Large \bf Partonic description of soft high energy pp interactions\footnote{Presented at Linear Collider 2011: Understanding QCD at Linear Colliders  in searching for old and new physics, 12-16 September 2011, ECT*, Trento, Italy}}

\vspace*{1cm}
                                                   
A.D. Martin$^a$, V.A. Khoze$^{a,b}$ and M.G. Ryskin$^{a,b}$  \\                                                    
                                                   
\vspace*{0.5cm}                                                    
$^a$ Institute for Particle Physics Phenomenology, University of Durham, Durham, DH1 3LE \\                                                   
$^b$ Petersburg Nuclear Physics Institute, Gatchina, St.~Petersburg, 188300, Russia \\          
                                                    
\vspace*{1cm}                                                    
                                                    
\begin{abstract}                                                    
We discuss how the main features of high-energy `soft' and `semihard' $pp$ collisions may be described in terms of parton cascades and multi-Pomeron exchange. The interaction between Pomerons produces an effective infrared cutoff, $k_{\rm sat}$, by the absorption of low $k_t$ partons. This provides the possibility of extending the parton approach, used for `hard' processes, to also describe high-energy soft and semihard interactions. We outline a model which incorporates these features. Finally, we discuss what the most recent LHC measurements in the soft domain imply for the model.

\end{abstract}                                                        
\vspace*{0.5cm}                                                    
                                                    
\end{center}

\section{A unified description?}

`Soft' and `hard' high-energy $pp$ interactions are described in different ways.  The appropriate formalism for high-energy soft interactions is based on Reggeon Field Theory with a phenomenological (soft) Pomeron, whereas for hard interactions we use a QCD partonic approach, where the (QCD) Pomeron is associated with the BFKL vacuum singularity \cite{book}. However, the two approaches appear to merge naturally into one another.  That is, the partonic approach seems to extend smoothly into the soft domain. 

\begin{figure}[htb]
\begin{center}
\resizebox{0.9\textwidth}{!}{
\includegraphics{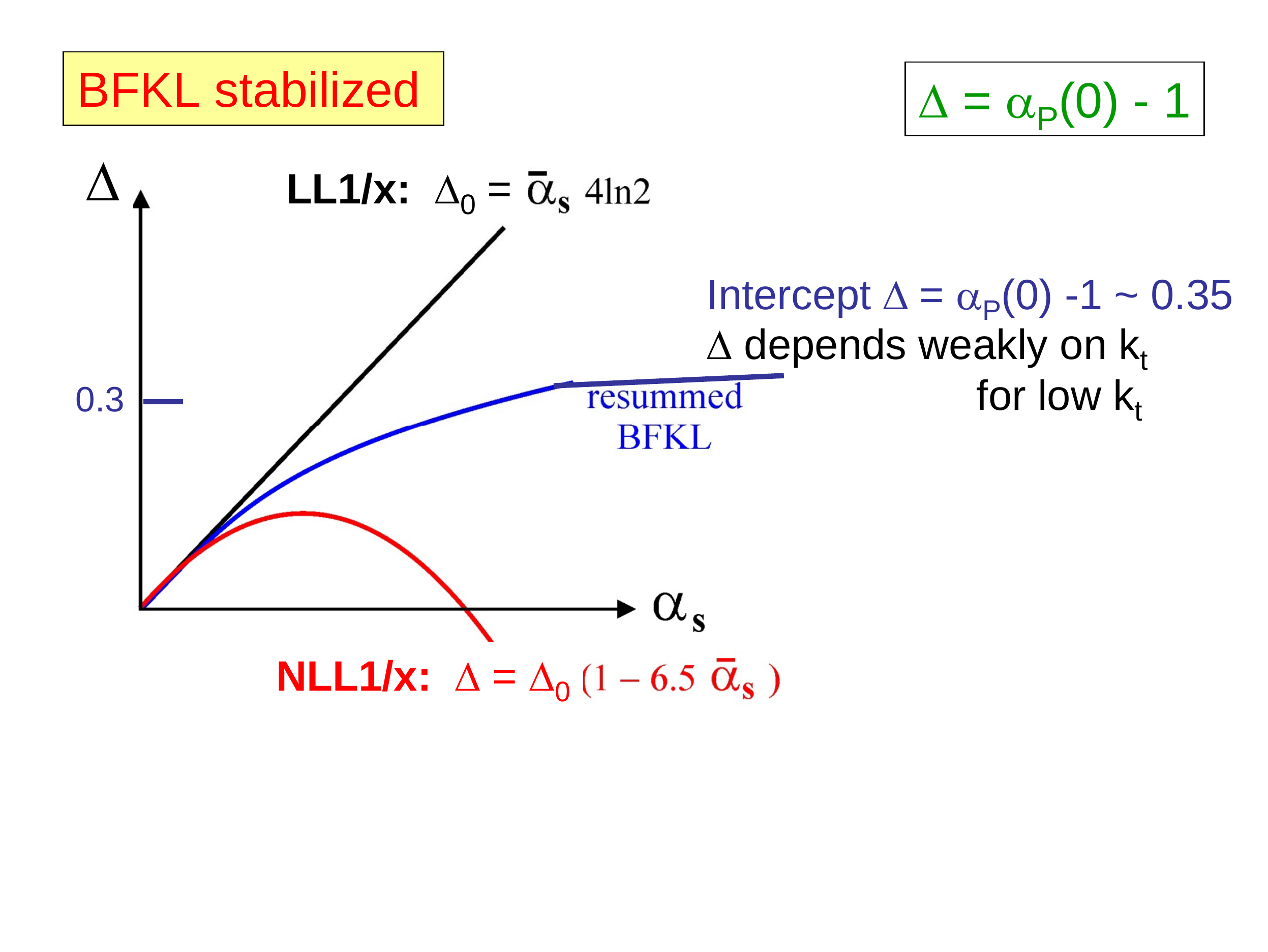}\hspace{1cm}
}
\vspace{-3cm}
\caption{\sf The behaviour found for the Pomeron intercept at leading and next-leading log$(1/x)$ order, where $\bar{\alpha}_s\equiv\alpha_s/3\pi$. When an all-order resummation of the main high-order contributions is included, $\Delta$ tends to a value of about 0.35 for reasonably large values of $\alpha_s$.}
\label{fig:BFKLstab}
\vspace{-0.5cm}
\end{center}
\end{figure}

The BFKL equation describes the development of the gluon shower as the momentum fraction, $x$, of the proton carried by the gluon decreases.  That is, the evolution parameter is ln$(1/x)$, rather than the ln$k_t^2$ evolution of the DGLAP equation. 
Formally, to justify the use of perturbative QCD, the BFKL equation should be written for gluons with sufficiently large $k_t$. However, it turns out that, after accounting for NLL$(1/x)$ corrections and performing an all-order resummation of the main higher-order contributions \cite{bfklresum}, the intercept of the BFKL Pomeron depends only weakly on the scale for reasonably small scales. The intercept is seen to be $\Delta \equiv \alpha_P(0)-1 \sim 0.35$ over a large interval of smallish $k_t$, Fig.~\ref{fig:BFKLstab,kmrs}.
Thus the BFKL Pomeron is a natural object to continue from the `hard' domain into the `soft' region.
\begin{figure}[htb]
\begin{center}
\vspace{-3.1cm}
\resizebox{0.8\textwidth}{!}{
\includegraphics{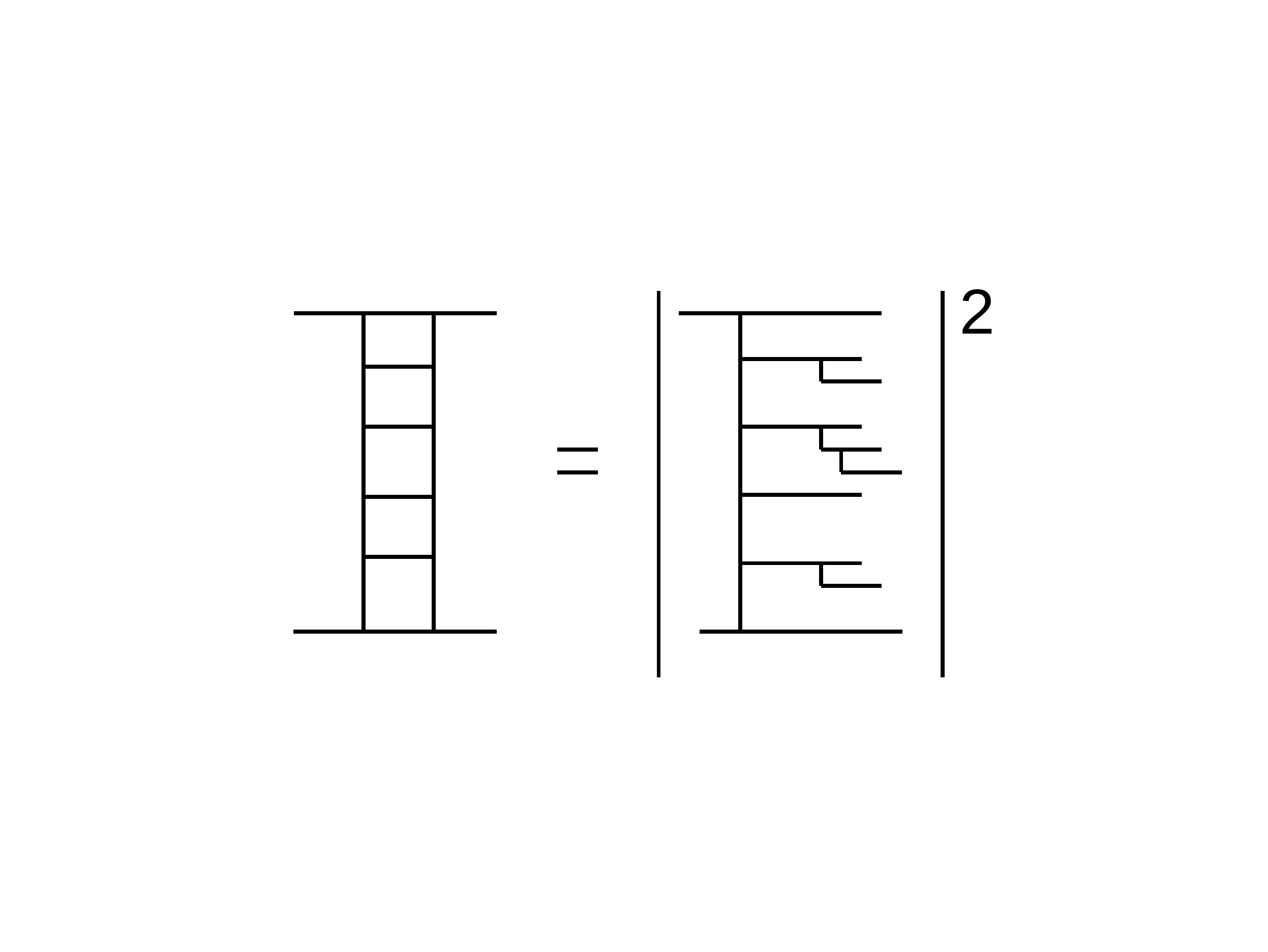}\hspace{1cm}
}
\vspace{-3cm}
\caption{\sf The cascade structure of a gluon ladder. The BFKL or QCD Pomeron is the sum of ladder diagrams, each with a different number of rungs.}
\label{fig:cascade}
\vspace{0.1cm}
\end{center}
\end{figure}

The BFKL or QCD Pomeron may be viewed as a sum of ladders based on the exchange of two $t$-channel (Reggeized) gluons. Each ladder produces a gluon cascade which develops in ln$(1/x)$ space, and which is not strongly ordered in $k_t$, see Fig.~\ref{fig:cascade}. There are phenomenological arguments (such as the small slope of the Pomeron trajectory\footnote{Recall that $\alpha'_P \propto 1/\langle k_t^2 \rangle \propto R_{\rm Pom}^2$.}, the success of the Additive Quark Model relations\footnote{The argument is best seen by analogy with nuclear physics. For light nuclei we have  `additive' cross sections, $\sigma=A_1A_2 \sigma_{nn}$, since the nuclei radii $R \gg r_{nn}$. On the other hand for a heavy nucleus, where $r_{nn}\sim R$, large Glauber corrections break the additive result. Similarly, the experimental success of the AQM indicates that $r_{qq}\sim R_{\rm Pom} \ll R_p$.}, etc.) which indicate that the size of an individual Pomeron is relatively small as compared to the size of a proton or pion etc. Thus we may regard the cascade as a small-size `hot-spot' inside the colliding protons.

At LHC energies the interval of BFKL ln$(1/x)$ evolution is much larger than that for DGLAP ln$k_t^2$ evolution. Moreover, the data already give hints that we need contributions not ordered in $k_t$, $\grave{a}~ la$ BFKL, since typically DGLAP overestimates the observed $\langle k_t \rangle$ and underestimates the mean multiplicity \cite{cms,atlas}. Further,  it is not enough to have only one Pomeron ladder exchanged; we need to include multi-Pomeron exchanges.

\begin{figure}[htb]
\begin{center}
\resizebox{1.0\textwidth}{!}{
\includegraphics{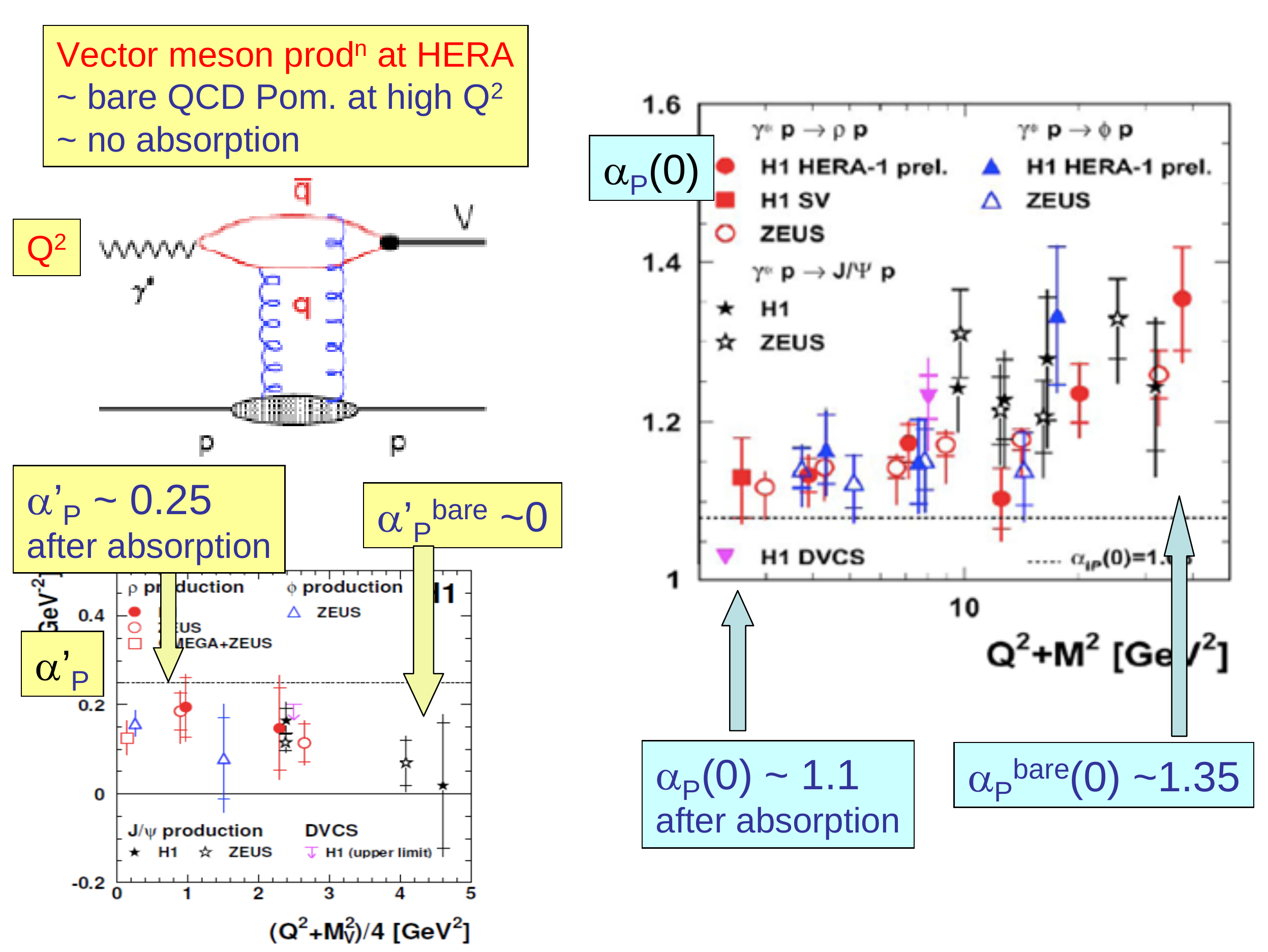}\hspace{1cm}
}
\caption{\sf The parameters of the Pomeron trajectory, $\alpha_P(t)=\alpha_P(0)+\alpha'_Pt$, determined from the energy and $t$ behaviour of high energy HERA data for vector meson production, $\gamma^*p\to V(M)p$.}
\label{fig:VM}
\vspace{0.1cm}
\end{center}
\end{figure}

Basically, the picture is as follows. In the perturbative domain we have a single bare `hard' Pomeron exchanged with a trajectory $\alpha_P^{\rm bare}\simeq 1.35+\alpha'_{\rm bare}t$, where $\alpha'_{\rm bare} \lapproxeq 0.05$ GeV$^{-2}$. The transition to the soft region is accompanied by absorptive multi-Pomeron effects, such that an {\it effective} `soft' Pomeron may be approximated by a linear trajectory $\alpha^{\rm eff}_P \simeq 1.08+0.25t$ in the {\it limited} energy range up to Tevatron energies \cite{DL}.  This smooth transition from hard to soft  is well illustrated by Fig.~\ref{fig:VM}, which shows 
 the behaviour of the data for vector meson ($V=\rho, \omega, \phi, J/\psi$) production at HERA, $\gamma^*p\to V(M)p$, as $Q^2+M^2$ decreases from about 50 GeV$^2$ towards zero.

\section{Multi-Pomeron diagrams}

The {\it eikonal} model accounts for the multiple rescattering of the incoming fast particles. We have\footnote{To allow for low-mass proton dissociation, the amplitude (\ref{eq:eik}) is written in matrix form, $T_{ik}$, between (Good-Walker \cite{GW}) diffractive eigenstates.}
\begin{equation}
{\rm Im}T\;=\;(1-e^{-\Omega/2})\;=\;(\Omega/2)-(\Omega^2/8)+...
\label{eq:eik}
\end{equation} 
which displays the multi-Pomeron corrections to the bare Pomeron amplitude, $\Omega/2$, that
 tame the power 
growth of the cross section with energy. Simultaneously, 
these multi-Pomeron diagrams also explain the growth of the central 
plateau
\cite{cms,atlas}
\be
\frac{dN}{d\eta}~=~n_P \frac{dN_{\rm 1-Pom}}{d\eta},
\label{eq:nP}
\ee
where $dN_{\rm 1-Pom}/d\eta$ is the plateau due to the exchange of one 
Pomeron, which is independent of collider energy. The growth is due to the 
increasing number, $n_P$, of Pomerons exchanged as energy increases. These 
(eikonal) multi-Pomeron contributions are included in the present Monte 
Carlos to some extent, as a Multiple Interaction (MI)
option, but Pomeron-Pomeron interactions are not allowed for. 

Since the (small size) Pomeron cascades (hot spots) occur at 
different impact parameters, $b$, there is practically no interference between them. Moreover, at this `eikonal' stage, 
the multi-Pomeron vertices, which account for the interaction between 
Pomerons, are not yet included in the formalism. These are interactions between partons within an individual hot spot (Pomeron). Formally, these are NNLO interactions, but their contribution is {\it enhanced} by the large multiplicity of partons within a high-energy cascade.  In terms of Reggeon Field Theory, the additional interactions are described by so-called {\it enhanced} multi-Pomeron diagrams, whose contributions are controlled by triple-Pomeron (and more complicated multi-Pomeron) couplings\footnote{These diagrams are responsible for high-mass proton dissociation.}. Recall that non-enhanced (eikonal) multi-Pomeron interactions are caused mainly by Pomerons occurring at different impact parameters, and well separated from each other in the $b$-plane. On the other hand, the enhanced contributions mainly correspond to additional interactions (absorption) within an {\it individual} hot spot, but with the partons well separated in rapidity.

The main effect of the enhanced contribution is the absorption of low $k_t$ partons. Note that the probability of these additional interactions is proportional to $\sigma_{\rm abs} \sim 1/k_t^2$, and their main qualitative effect is to induce a splitting of low $k_t$ partons into a pair of partons each with lower $x$, but larger $k_t$. Effectively this produces a dynamical infrared cut-off, $k_{\rm sat}$, on $k_t$, and partly restores a DGLAP-like $k_t$-ordering within the cascade at larger $k_t$.

\section{Schematic sketches of the model}

\begin{figure}[htb]
\begin{center}
\vspace{-3.5cm}
\resizebox{1.0\textwidth}{!}{
\includegraphics{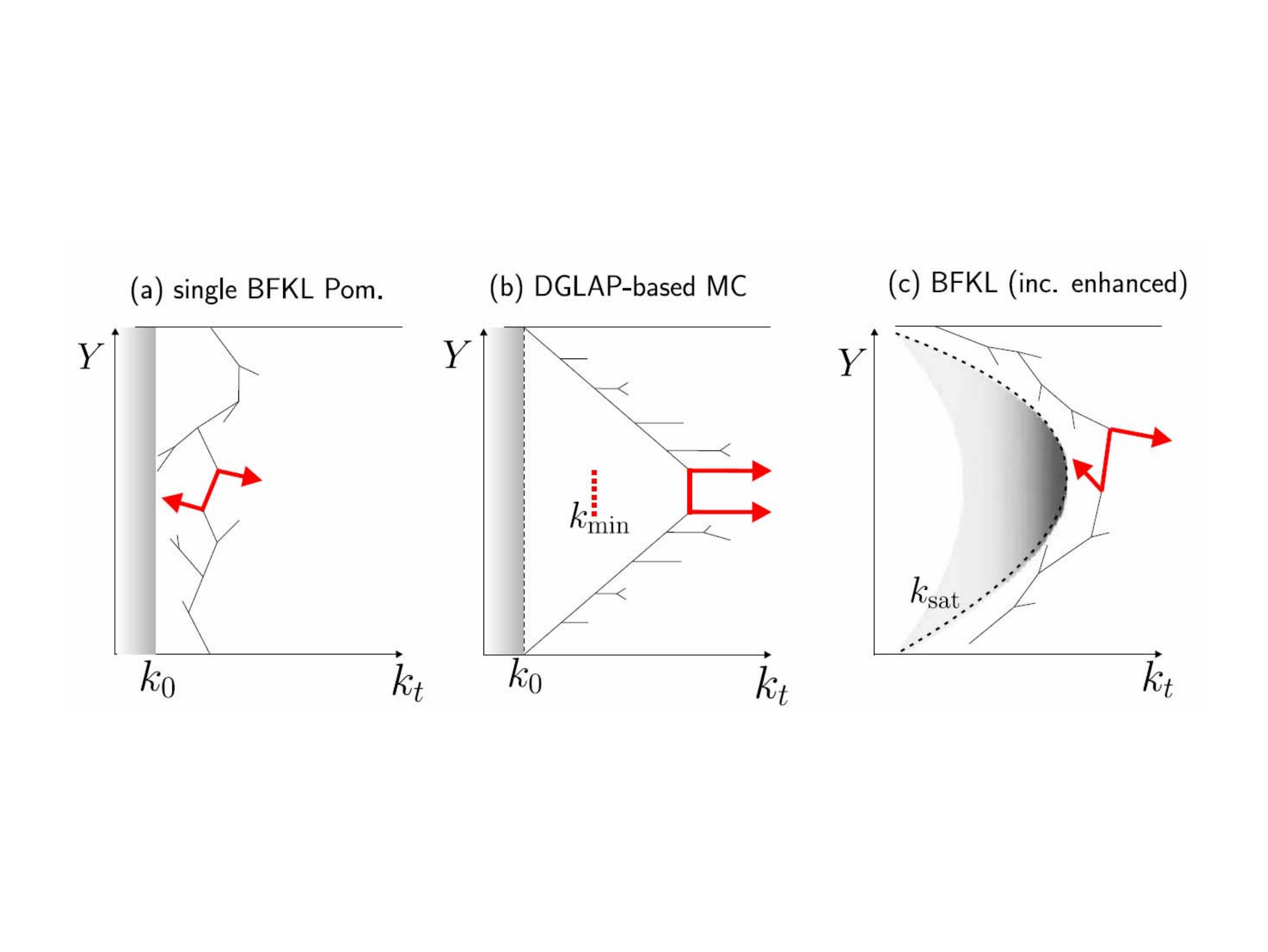}\hspace{1cm}
}
\vspace{-3.5cm}
\caption{\sf Sketches of the basic diagram for semi-hard particle production in $pp$ collisions. The figure is taken from Ref. \cite{KMRmp}.}
\label{fig:abc}
\end{center}
\end{figure}

Qualitatively, the structure of soft interactions based on the `BFKL' multi-Pomeron approach is as follows. The evolution 
 produces a parton cascade which occupies a relatively small domain in $b$-space, as compared to the size of the proton. We have called this a hot spot. The multiplicity of partons grows as $x^{-\Delta}$, while the $k_t$'s of the partons are not strongly ordered and depend weakly on ln$s$. Recall $\Delta\equiv \alpha_P(0)-1$. Allowing for the running of $\alpha_s$, the partons tend to drift to lower $k_t$ where the coupling is larger. This is shown schematically in Fig.~\ref{fig:abc}(a). 

On the contrary, the DGLAP-based Monte Carlos generate parton cascades strongly ordered in $k_t$. That is, the parton $k_t$ increases as we evolve from the input PDF of the proton to the matrix element of the hard subprocess, which occurs near the centre of the rapidity interval, Fig.~\ref{fig:abc}(b).  Since the cross section of the hard subprocess behaves as $d\hat{\sigma}/dk_t^2 \propto 1/k_t^4$, the dominant contributions come from near the lower limit $k_{\rm min}$, of the $k_t$ integration. In  fact, in order to describe the high-energy collider data, it is necessary to artificially introduce an energy dependent infrared cutoff; $k_{\rm min} \propto s^a$ with\footnote{This value of $a$ is very close to that obtained from the resummed NLL
BFKL prediction that the saturation scale satisfies $Q^2_{\rm sat}(x)\propto x^{-0.45}$ \cite{kmrs}. Since $x\propto 1/\sqrt{s}$ and $Q_{\rm sat}=k_{\rm min}$,
 we expect $a=0.45/4$. } $a \sim 0.12$ \cite{P81}. This cutoff is only applied to the hard matrix element, whereas in the evolution of the parton cascade a constant cutoff $k_0$, corresponding to the input PDFs, is used. Note that during the DGLAP evolution, the position of the partons in $b$-space is frozen. Thus such a cascade also forms a hot spot.

Accounting for  the multiple interaction option, that is for contributions containing a few hot spots, we include the eikonal multi-Pomeron contributions, both for the DGLAP and BFKL based descriptions; that is the presence of a few small-size QCD Pomeron cascades.

Next, we include the enhanced multi-Pomeron diagrams  
introducing the absorption of the low $k_t$ partons. The strength of absorption is driven by the parton density and therefore the effect grows with energy, that is with ln$(1/x)$. We thus have an effective infrared cutoff, $k_{\rm sat}(x)$, which modifies the $k_t$ distribution of the `BFKL' cascade. The result is shown Fig.~\ref{fig:abc}(c), which has some similarity to the DGLAP cascade of Fig.~\ref{fig:abc}(b). However, now the cutoff $k_{\rm sat}$ is not a tuning parameter, but is generated dynamically by the enhanced multi-Pomeron diagrams. Recall that the same diagrams describe high-mass proton dissociation.  That is, the value of the multi-Pomeron vertex simultaneously controls the cross sections of high-mass dissociation and the effective cutoff $k_{\rm sat}$ -- two phenomena which, at first sight, appear to be quite different.

\section{The Durham model}
\begin{figure}[htb]
\begin{center}
\vspace{-4cm}
\resizebox{0.9\textwidth}{!}{
\includegraphics{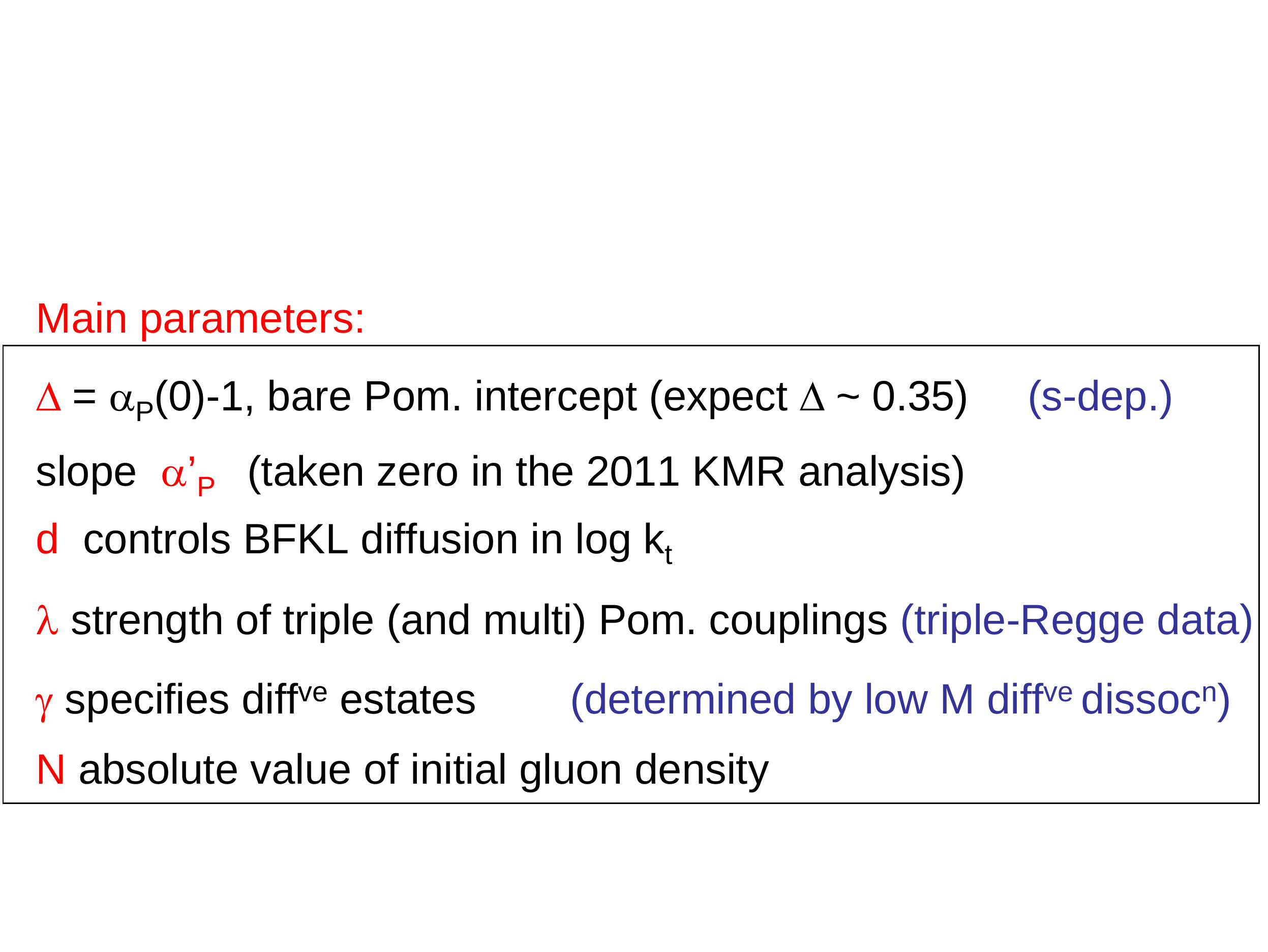}\hspace{1cm}
}
\vspace{-1.5cm}
\caption{\sf Some of the main parameters of the KMR model \cite{KMRnnn}; and how they are constrained.}
\label{fig:parameters}
\vspace{-0.5cm}
\end{center}
\end{figure}

How may the partonic model of the Pomeron be implemented in practice? To achieve this we note that  the absorption of low $k_t$ partons is 
driven by the opacity, $\Omega$, which depends both on $k_t$ and 
$y=\ln(1/x)$. The opacity, $\Omega_{ik}(y,k_t,b)$, is obtained \cite{KMRnnn} by solving the corresponding BFKL-type
evolution equations in $y$ with a simplified form of the kernel, but which incorporates the main features of BFKL: diffusion in ln$k_t^2$ and $\Delta=\alpha_P^{\rm bare}(0)-1\simeq 0.35$. (A two-channel eikonal is used, $i,k=1,2$.) The inclusion of the $k_t$ dependence is crucial for the transition from the hard to the soft domain. The absorptive factors in the equation embody the result that there is less screening for larger $k_t$. 
The model \cite{KMRnnn} has only a small number of physically motivated parameters, see Fig.~\ref{fig:parameters}. whose values are tuned to reproduce the available high energy $pp$ and $p\bar{p}$ data for $\sigma_{\rm tot}, \;d\sigma_{\rm el}/dt,\; \sigma_{\rm SD}^{{\rm low}M}, \;\sigma_{\rm SD}^{{\rm high}M}/dtdM^2$ etc. Given  $\Omega_{ik}(y,k_t,b)$ we can, in principle, predict all soft and semi-hard inclusive phenomena, such as the survival factors of rapidity gaps, the PDFs and diffractive PDFs at low $x$ and low scales, etc.  The predictions for some of the cross sections are given in the left-half of Table \ref{tab:A2}, as their values will be relevant for the discussion in Section \ref{sec:5}.
\begin{table}[htb]
\begin{center}
\begin{tabular}{|c||cccc||ccccc||}\hline
 &    &  KMR & model& & &KMR & 3-ch &eikonal  &\\ \hline
energy &   $\sigma_{\rm tot}$ &  $\sigma_{\rm el}$  &  $\sigma^{\rm SD}_{{\rm low}M}$ & $\sigma^{\rm DD}_{{\rm low}M}$ & $\sigma_{\rm tot}$ &  $\sigma_{\rm el}$ & $B_{\rm el}$ &  $\sigma^{\rm SD}_{{\rm low}M}$ & $\sigma^{\rm DD}_{{\rm low}M}$ \\ \hline

 1.8  & 72.7 & 16.6&4.8& 0.4 &  79.3  &     17.9  &       18.0      &   5.9 &0.7   \\
 7   & 87.9 & 21.8&   6.1& 0.6 & 97.4    &    23.8    &     20.3  &      7.3 & 0.9    \\
14    & 96.5 & 24.7 & 7.8 &0.8 & 107.5  &     27.2     &  21.6   &    8.1 & 1.1  \\
 100   & 122.3 & 33.5 & 9.0 & 1.3 & 138.8  &     38.1  &  25.8  &  10.4  & 1.6     \\
 \hline

\end{tabular}
\end{center}
\caption{\sf Some results of the complete KMR model \cite{KMRnnn} prior to the LHC data (left-hand Table), and results obtained from a simpler approach, described in Section \ref{sec:5}, based on a 3-channel eikonal description \cite{KMR12} of all elastic (and quasi-elastic) $pp$ and $p\bar{p}$ data, including the TOTEM LHC data (right-half of the Table). $\sigma_{\rm tot}$,  $\sigma_{\rm el}$ and    $\sigma^{\rm SD,DD}_{{\rm low}M}$ are the total, elastic and low-mass single and double dissociation cross sections (in mb) respectively.  The cross section $\sigma^{\rm SD}$ is the sum of the dissociations of both the `beam' and `target' protons. $B_{\rm el}$ is the mean elastic slope (in $\GeV^{-2}$), $d\sigma_{\rm el}/dt=e^{B_{\rm el}t}$, in the region $|t|<0.2~\GeV^2$.  The collider energies are given in TeV.  The former (latter) analysis fit to the CERN-ISR observations that $\sigma^{\rm SD}_{{\rm low}M}$=2(3) mb at $\sqrt{s}=53$ GeV, with low mass defined to be $M<2.5(3)$ GeV.}
\label{tab:A2}
\end{table}
It is important to note that hadronization can be incorporated in this partonic description of the Pomeron, via 
Monte Carlo generators, which now would have the advantage of an effective dynamical cutoff $k_{\rm sat}$ to suppress low $k_t$ parton emissions.

In summary, some of the main features of the model are:

(i) values of the high energy $pp$ total cross section which are suppressed by absorptive corrections. Increasingly large values of $\sigma^{{\rm high}M}_{\rm SD}$ are found due to the increasing phase space with collider energy.

\begin{figure}[htb]
\begin{center}
\vspace{-0.5cm}
\resizebox{0.9\textwidth}{!}{
\includegraphics{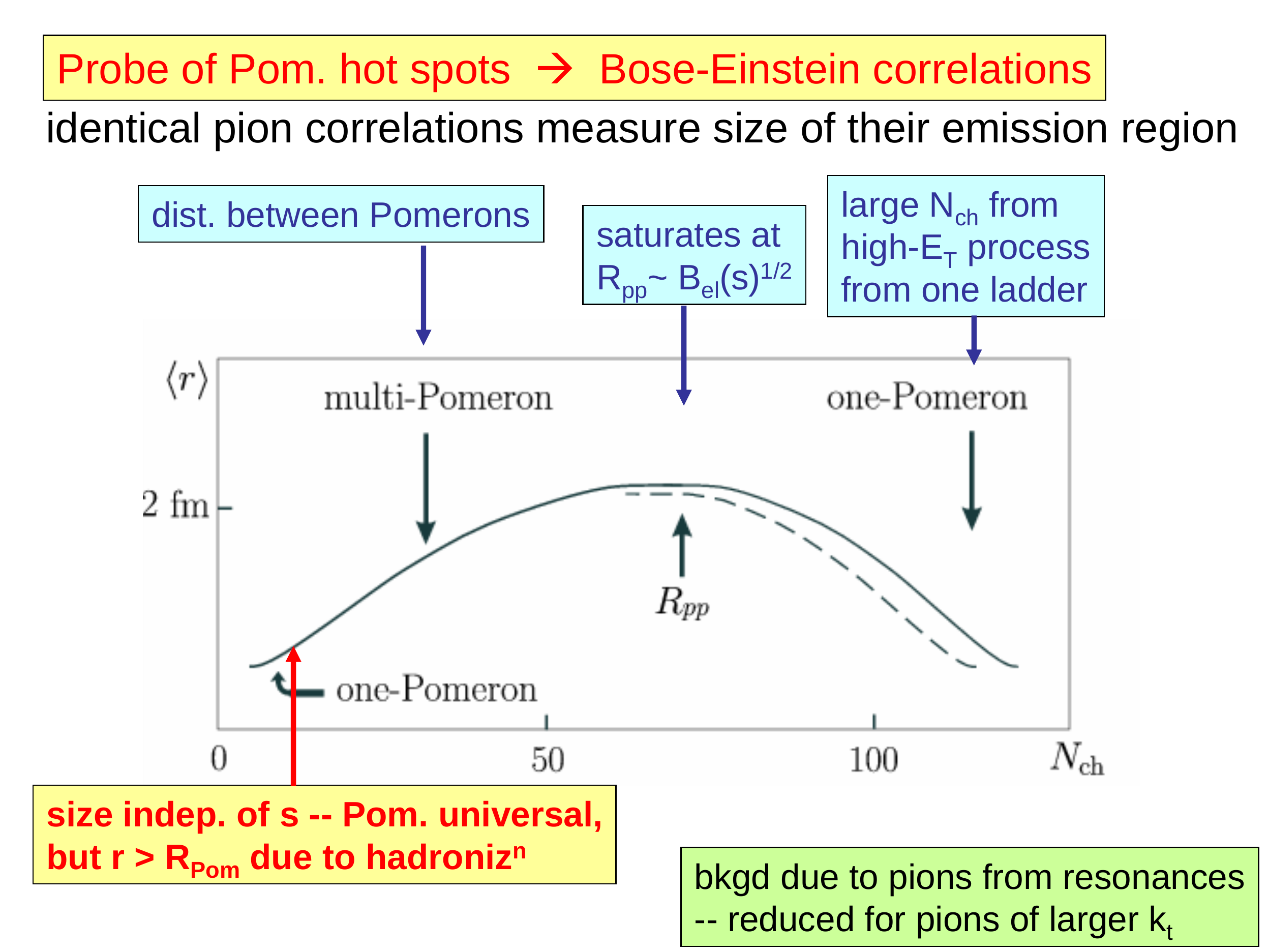}\hspace{1cm}
}
\caption{\sf A sketch of the size $\langle r \rangle$ of the source of identical pions as a function of the multiplicity of charged particles in high-energy $pp$ interactions \cite{BEC}. The continuous and dashed curves correspond to, say, $\sqrt{s}=7$ and 0.9 TeV respectively. At low $N_{\rm ch}$ we expect $\langle r \rangle$ to be independent of collider energy, $\sqrt{s}$, while for the plateau we expect $\langle r \rangle \sim R_{pp} \propto \sqrt{B_{\rm el}(s)}$ to increase very slowly with energy. Very high multiplicities are expected to arise from high-$E_T$ events originating from a single ladder.}
\label{fig:BEC}
\vspace{-0.1cm}
\end{center}
\end{figure}

(ii) multi-Pomeron contributions arising from eikonal diagrams, that is the presence of a few small-size QCD Pomeron cascades (hot spots). This can be tested by measuring Bose-Einstein correlations, see Fig.~\ref{fig:BEC}. Specifically, identical pion correlations measure the size of their emission region.

(iii)  multi-Pomeron contributions arising from enhanced diagrams, which lead to the absorption of low $k_t$ partons and automatically introduce an effective cutoff $k_{\rm sat}$ which increases with energy. Due to the cutoff, $k_t>k_{\rm sat}$, the main inelastic process is {\it minijet} production. The dominance of minijets can be tested by observing the two-particle correlations of secondaries at the LHC \cite{KMRmp}.

\section{Implications of latest LHC `soft' data \label{sec:5}}

As a postscript to the Durham approach, we briefly discuss the implications of some recent LHC data on `soft' diffractive processes which became available after the LC11 meeting.

First, we look at the implications of the recent TOTEM measurements, at 7 TeV, of $d\sigma_{\rm el}/dt$ down to $-t=0.02 ~\GeV^{-2}$ \cite{TOTEM2}. From these data, TOTEM find
\begin{equation}
\sigma_{\rm tot}=98.3 ~ {\rm mb},~~~~~~\sigma_{\rm el}=24.8~{\rm mb},~~~~~~\sigma_{\rm inel}=73.5~{\rm mb}.
\label{eq:T}
\end{equation}
In the discussion below, we will ignore the (important) experimental errors, just to get some ideas of the trends of the data. The KMR model \cite{KMRnnn} predicts lower values of 88, 22 and 66 mb respectively, see the left-half of Table \ref{tab:A2}. The model was tuned to describe collider data for $\sigma_{\rm tot}$. At the Tevatron energy, where the CDF \cite{CDF} and E710 \cite{E710} measurements disagree by some 10$\%$, we were much closer to the lower E710 value.

\begin{figure} 
\begin{center}
\includegraphics[height=18cm]{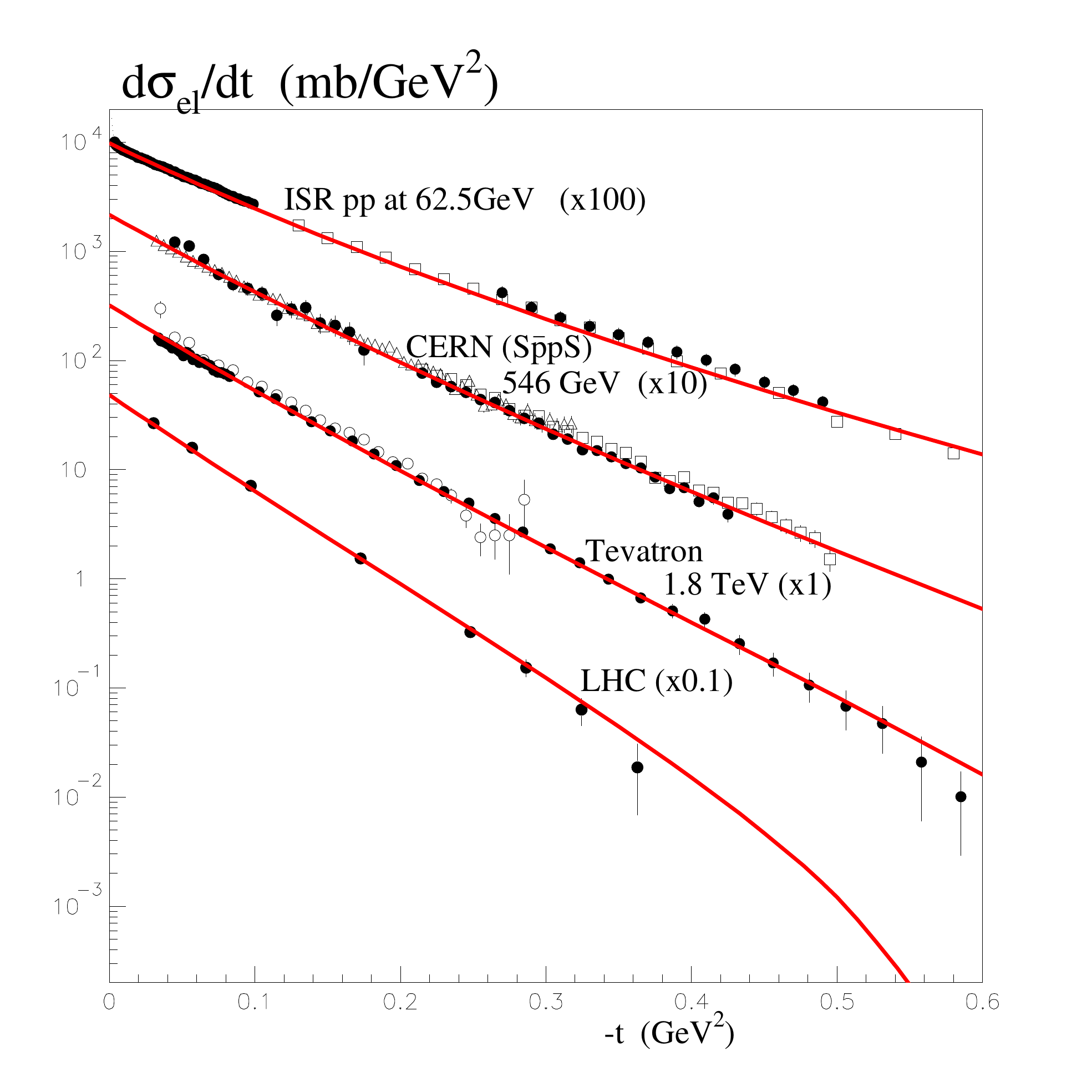}
\vspace{-0.8cm}
\caption{\sf The description of the data for the differential cross sections for $pp$ (or $p\bar{p}$) elastic scattering in the energy range 62.5 to 7000 GeV \cite{TOTEM2, SppS, CDF, E710, ISR} using a 3-channel eikonal model. The Tevatron data with open and closed circles are those of the CDF and E710 collaborations respectively. Only very selected TOTEM points are shown, which have been read off their published plot. The excellent agreement of our model with the data for small $|t|$ is achieved with a very economical parametrization of the diffractive eigenstates. It is straightforward to describe the elastic data in the region of the diffractive LHC dip \cite{TOTEM1}, but at the expense of a more complicated parametrization of the form factors of the three eigenstates.}
\label{fig:elastic}
\end{center}
\end{figure}

To investigate this further, we performed a simpler study than that in \cite{KMRnnn}. The idea was to see if we can describe all the elastic $pp$ and $p\bar{p}$ collider data in terms of a 3-channel eikonal model with only one Pomeron, with parameters that are naturally linked to the perturbative QCD (BFKL) framework, as discussed in the previous sections of this paper. However, for the simpler study \cite{KMR12}, we used an effective Pomeron, rather than the bare QCD Pomeron with intercept $\Delta_{\rm bare} \equiv \alpha_P(0)-1=0.32$ of \cite{KMRnnn}. With an economical parametrization of the three (Good-Walker) diffractive eigenstates, we are, indeed, able to obtain a good description of all these data for $|t|\lapproxeq 0.3~\GeV^2$ with $\Delta_{\rm eff}=0.14$, see Fig. \ref{fig:elastic}. Since our eikonal model was devised to fit the data it is not surprising to have agreement with the TOTEM cross sections of (\ref{eq:T}). We call the Pomeron `effective' since, although we accounted for eikonal rescattering of the incoming partons, unlike \cite{KMRnnn}, we did not explicitly consider enhanced rescattering involving intermediate partons. The latter are included implicitly since their main effect is to renormalize the bare Pomeron trajectory.

However, some observations can be made. First, the eikonal description is close to the CDF total cross section value, and yet the description of the E710 $d\sigma_{\rm el}/dt$ measurements is good.  Secondly, the insertion of the pion loop\footnote{
Recall that the pion loop insertion modifies the Pomeron trajectory at very small $t$ \cite{Anselm}. Indeed the presence of the $2\pi$ singularity at $t=4 m_{\pi}^2$ leads to some curvature in the $t$ behaviour of $d\sigma_{\rm el}/dt$. That is, to some variation of the local elastic slope $B_{\rm el}(t)$. Including the pion loop gives an equally good description of the elastic data.}  into the Pomeron trajectory would decrease the total cross value in Table \ref{tab:A2} by 1 mb, to 96.4 mb \cite{KMR12}.  The conclusion is that there is quite a bit of uncertainty in the extrapolation of the $d\sigma_{\rm el}/dt$ data to the optical point, in addition to the 3-4$\%$ normalization uncertainty. Future precise elastic measurements\footnote{Note also that the simultaneous  measurement of  bremsstrahlung
photons, 
accompanying elastic proton-proton scattering in CMS will, with the help
of the Zero Degree Calorimeter, allow an independent determination of $\sigma_{\rm el}/\langle B_{\rm el} \rangle$; see \cite{brems} for details.} 
even closer to $t=0$ will help reduce the uncertainty in the value of $\sigma_{\rm tot}$.  If the values of $\sigma_{\rm tot}$ and $d\sigma_{\rm el}/dt$ at the LHC are confirmed to be significantly higher than those obtained in \cite{KMRnnn}, then this full analysis should be repeated with these data included. It will result in a somewhat larger value of $\Delta_{\rm bare}$.

Let us now compare the results shown in Table~\ref{tab:A2} with the inelastic cross section obtained by CMS, ATLAS and ALICE at 7 TeV. The measured value is defined as the cross section with at least two particles in some central (but far from complete) rapidity, $\eta$, interval.  For instance, ATLAS find 
$\sigma_{\rm inel}=60.3$ mb for the cross section of processes with $M>15.7$ GeV, that is $\xi =M^2/s >5 \times 10^{-6}$ \cite{ATLAS}.  After a model dependent extrapolation to cover the entire rapidity interval they obtain $\sigma_{\rm inel}=69.4$ mb.  CMS find a very similar result, namely 68.0 mb \cite{CMS1}. ALICE also get a similar result \cite{ALICE}.  
These estimates are about 5 mb lower than the recent TOTEM value of 73.5 mb of ({\ref{eq:T}).
The difference may be attributed to the extrapolated values being 5 mb deficient for low-mass diffraction. (The extrapolation in the high-mass interval is confirmed by the ATLAS measurement $d\sigma/d\Delta\eta \simeq d\sigma/d{\rm ln} M^2 \simeq 1$ mb per unit of rapidity \cite{Atlas}.) More specifically, if we define low mass to be $M<3$ GeV, then, noting that the unmeasured interval from $M=15.7$ to $M=3$ GeV gives $\Delta{\rm ln} M^2 =3.3  $, it follows that the ATLAS, CMS results imply $\sigma_{\rm inel}^{{\rm high}M}\simeq 64$ mb. Then using the TOTEM result we find that low-mass diffractive dissociation is expected to have a rather large cross section
\begin{equation}
\sigma_{\rm inel}^{{\rm low}M}\simeq 73.5-64 ~=~ 9.5 \;{\rm mb}.
\label{eq:low}
\end{equation}
Note, however, that the low-mass diffractive dissociation given in Table \ref{tab:A2},
\begin{equation}
\sigma^{\rm SD+DD}_{{\rm low}M}~=~7.3+0.9~=~8.2 ~{\rm mb},
\end{equation} 
in satisfactory agreement with (\ref{eq:low}).

Here, it is worth noting that, as seen in \cite{minbias}, by triggering inelastic events
with the T1 and T2 forward tracking telescopes \cite{TOTEM3},
TOTEM can obtain an independent estimate  of
the total inelastic cross section, analogous to that of the
ATLAS measurement \cite{ATLAS} of $\sigma_{\rm inel}$, which used an
information from
the MBTS scintillation counters.
In particular, note that since the T2 telescope extends to
larger values of pseudorapidity ($5.3<|\eta|<6.5$), this allows
a wider coverage of high-mass diffraction (down to $M\sim 4 $ GeV).
This, in turn, may allow the extrapolation to the
total value of $\sigma_{\rm inel}$ to be performed with reduced uncertainty.

Another valuable set of soft diffractive measurements have been made by the ATLAS collaboration. They measure $d\sigma/d\Delta\eta$ versus $\Delta\eta$ for events with large rapdity gaps \cite{Atlas}. For $\Delta\eta \gapproxeq 5$, fluctuations in hadronization are greatly suppressed \cite{KKMRZ}, and we cleanly probe high-mass diffractive dissociation. In Ref. \cite{KMR12} these data are shown to be well described by a triple-Pomeron approach, {\it provided} the sizeable absorptive or rescattering corrections are taken into account. These corrections are computed in a parameter-free way using the 3-channel eikonal model discussed above.

\section*{Acknowledgements} We thank Giulia Pancheri for arranging such an enjoyable Workshop.


\begin{thebibliography}{99}
\bibitem{book} for a recent detailed review see
V.S.~Fadin, B.L.~Ioffe and L.N.~Lipatov,
{\it in} Quantum Chromodynamics
(Camb. Univ. Press, 2010).

\bibitem{bfklresum} M. Ciafaloni, D. Colferai and G. Salam, Phys. Rev. {\bf D60}, 114036 (1999).

\bibitem{kmrs} V.A.~Khoze, A.D.~Martin, M.G.~Ryskin and W.J. Stirling, Phys. Rev. {\bf
D70}, 074013 (2004).


\bibitem{cms} CMS Collaboration,
  Phys.\ Rev.\ Lett.\  {\bf 105}, 022002 (2010).
  
\bibitem{atlas}  ATLAS Collaboration, Phys. Rev. {\bf D83}, 112001 (2011). 
 

\bibitem{DL} A. Donnachie and P.V. Landshoff, Phys. Lett. {\bf B296}, 227 (1992).

\bibitem{GW} M.L. Good and W.D. Walker, Phys. Rev. {\bf 120}, 1857 (1960).

\bibitem{KMRmp} M.G. Ryskin, A.D. Martin and V.A. Khoze, J. Phys. G {\bf 38}, 085006 (2011).

\bibitem{P81} T.~Sjostrand, S.~Mrenna and P.Z.~Skands,
  Comput.\ Phys.\ Commun.\  {\bf 178}, 852 (2008).
   
\bibitem{KMRnnn} M.G. Ryskin, A.D. Martin and V.A. Khoze, Eur. Phys. J. {\bf C71}, 1617 (2011).

\bibitem{KMR12} M.G. Ryskin, A.D. Martin and V.A. Khoze, arXiv:1201.6298.

\bibitem{BEC} V.A.~Schegelsky, A.D.~Martin, M.G.~Ryskin and V.A.~Khoze, Phys. Lett. {\bf B703}, 288 (2011). 

\bibitem{TOTEM2} TOTEM Collaboration, Europhys. Lett. {\bf 96}, 21002 (2011).

\bibitem{CDF} CDF Collaboration, Phys. Rev. {\bf D50}, 5518 (1994).

 
\bibitem{E710} E710 Collaboration, Phys. Lett. {\bf B247}, 127 (1990).

\bibitem{SppS} UA4 Collaboration, Phys. Lett. {\bf B147}, 385 (1984);\\
UA4/2 Collaboration, Phys. Lett. {\bf B316}, 448 (1993); \\
UA1 Collaboration, Phys. Lett. {\bf B128}, 336 (1982).

\bibitem{ISR} N. Kwak et al., Phys. Lett. {\bf B58}, 233 (1975);\\
U. Amaldi et al., Phys. Lett. {\bf B66}, 390 (1977);\\
L. Baksay et al., Nucl. Phys. {\bf B141}, 1 (1978).


\bibitem{TOTEM1} TOTEM Collaboration, Europhys. Lett. {\bf 95}, 41001 (2011).

\bibitem{Anselm} A.A. Anselm and V.N. Gribov, Phys. Lett. {\bf B40}, 487 (1972);\\
V.A. Khoze, A.D. Martin and M.G. Ryskin, Eur. Phys. J. {\bf C18}, 167 (2000).

\bibitem{brems} V.A.~Khoze, J.W.~Lamsa, R.~Orava and M.G.~Ryskin,
  JINST {\bf 6}, P01005 (2011); \\
  H.~Gronqvist, V.A.~Khoze, J.W.~Lamsa, M.~Murray and R.~Orava,
  arXiv:1011.6141.


\bibitem{ATLAS} ATLAS Collaboration, Nature Commun. {\bf 2}, 463 (2011); arXiv:1104.0326 [hep-ex].

\bibitem{CMS1} CMS Collaboration, Note CMS-PAS-FWD-11-001, (2011).

\bibitem{ALICE} M.G.~Poghosyan, for the ALICE Collaboration, J. Phys. {\bf G38}, 124044 (2011). 

\bibitem{minbias} V.A.~Khoze, A.D.~Martin and M.G.~Ryskin,
  Phys.\ Lett.\ {\bf B679}, 56 (2009).

\bibitem{TOTEM3} TOTEM Collaboration, Nucl.\ Instrum.\ Meth.\  {\bf A617}, 62 (2010).


\bibitem{Atlas} ATLAS Collaboration, arXiv:1201.2808.

\bibitem{KKMRZ} V.A. Khoze et al., Eur. Phys. J. {\bf C69}, 85 (2010).


\end{thebibliography}
\end{document}